\documentclass[12pt,final]{iopart}

\usepackage{geometry}
\usepackage[parfill]{parskip}    % Activate to begin paragraphs with an empty line rather than an indent
\usepackage{graphicx}
\usepackage[ansinew]{inputenc}
\usepackage{palatino, url, multicol}
\usepackage[usenames,dvipsnames]{color}
\usepackage{iopams}  
\usepackage{subfigure}
\DeclareGraphicsExtensions{eps}
\usepackage[center]{caption}
\def \BeqA{\begin{eqnarray}}
\def \EeqA{\end{eqnarray}}

\def \Beq{\begin{equation}}
\def \Eeq{\end{equation}}

\begin{document}

\title[]{Birth and Death in Lineland\\
or\\
A Purely Geometric model of Creation and Annihilation of Particles in Low Dimension}

\author{T. Platini, R. Low}

\address{Applied Mathematics Research Center, Coventry University, Coventry CV1 5FB, England}

\eads{\mailto{thierry.platini@coventry.ac.uk}, \mailto{mtx014@coventry.ac.uk}}

\begin{abstract}
We consider a simple model of a one dimensional universe embedded in the Euclidean plane. In this model, a circle travelling at constant speed intersects the line in a pair of points which first separate and then rejoin. We interpret this as the creation and annihilation of a pair of particles, and describe the trajectories as seen by inhabitants of the one dimensional universe. In our final scenario, considering Lineland as curved, we show how sufficiently accurate measurements will push inhabitants of Lineland to reject their theory for a purely geometric interpretation.
\end{abstract}

%\pacs{PUT SOMETHING HERE}
\vspace{2pc}
\submitto{Eur J Phys}
\maketitle

%%%%%%%%%%%%%%%%%%%%%%%%%%%%%
%%%%%%%%%%%%%%%%%%%%%%%%%%%%%
\section{Introduction}
There is a long history of consideration of lower dimensional universes. The earliest was Edwin Abbott's \textit{Flatland} \cite{Abb1884}, but here although geometry was significant, its story serves the purposes of social satire. The most obvious descendant of this work is Dewdney's investigations of the planiverse \cite{Dewdney1983}, which makes a more serious study of physics and chemistry in two dimensional space. Less well known is Norton Juster's \textit{The Dot and the Line} \cite{Norton1963} set in a one (spatial) dimensional universe called Lineland, and which is again a social commentary.

In a more technical vein, we should mention that in quantum mechanics in particular, one-dimensional physics is of interest for experimental and theoretical reasons. In 1959, Loudon was first to consider 1D chemistry of the H atom \cite{Loudon1959}. His work had many applications in different scenarios where low dimensionality was key. Today, popular topics such as carbon nanotubes \cite{Saito1998}, Luttinger liquid \cite{Deshpande2010,Blumenstein2011,Laroche2014} and confined atomic gases \cite{Moritz2005,Guan2013}, push us to think in one dimension. In a recent paper, Loos and collaborators \cite{Loos2014} have studied 1D atomic and molecular systems of interacting particles to build the 1D Mendeleev's periodic table.

In our work we will investigate the physics emerging on a one-dimensional system (Lineland) as it is embedded in a 2D space. We will assume that Linelanders can only observe the points of intersection of a shape in the plane with their world. In particular we will consider the effective physics, appearing to inhabitants of Lineland, as circles on the 2D plane are passing through. To Linelanders this phenomenon appears as the creation of a pair of effective particles which separate, then turn about, approach each other and annihilate. We will see that the effective force between particles takes a particularly simple form. After considering the physics of particle pair creation and annihilation, we consider how the situation is affected if Lineland is curved. In particular we will investigate how the inhabitants of Lineland might discover that they in fact live in Circleland, and how they could estimate its radius, by sufficiently accurate measurements. Finally, we address the inverse problem: by looking for shapes which, when crossing Lineland, leads to a given effective force.

%%%%%%%%%%%%%%%%%%%%%%%%%
%%%%%%%%%%%%%%%%%%%%%%%%%
\begin{figure}
\centering
\begin{minipage}{.5\textwidth}
  \centering
    \label{schema1}
  \includegraphics[width=1.1\linewidth]{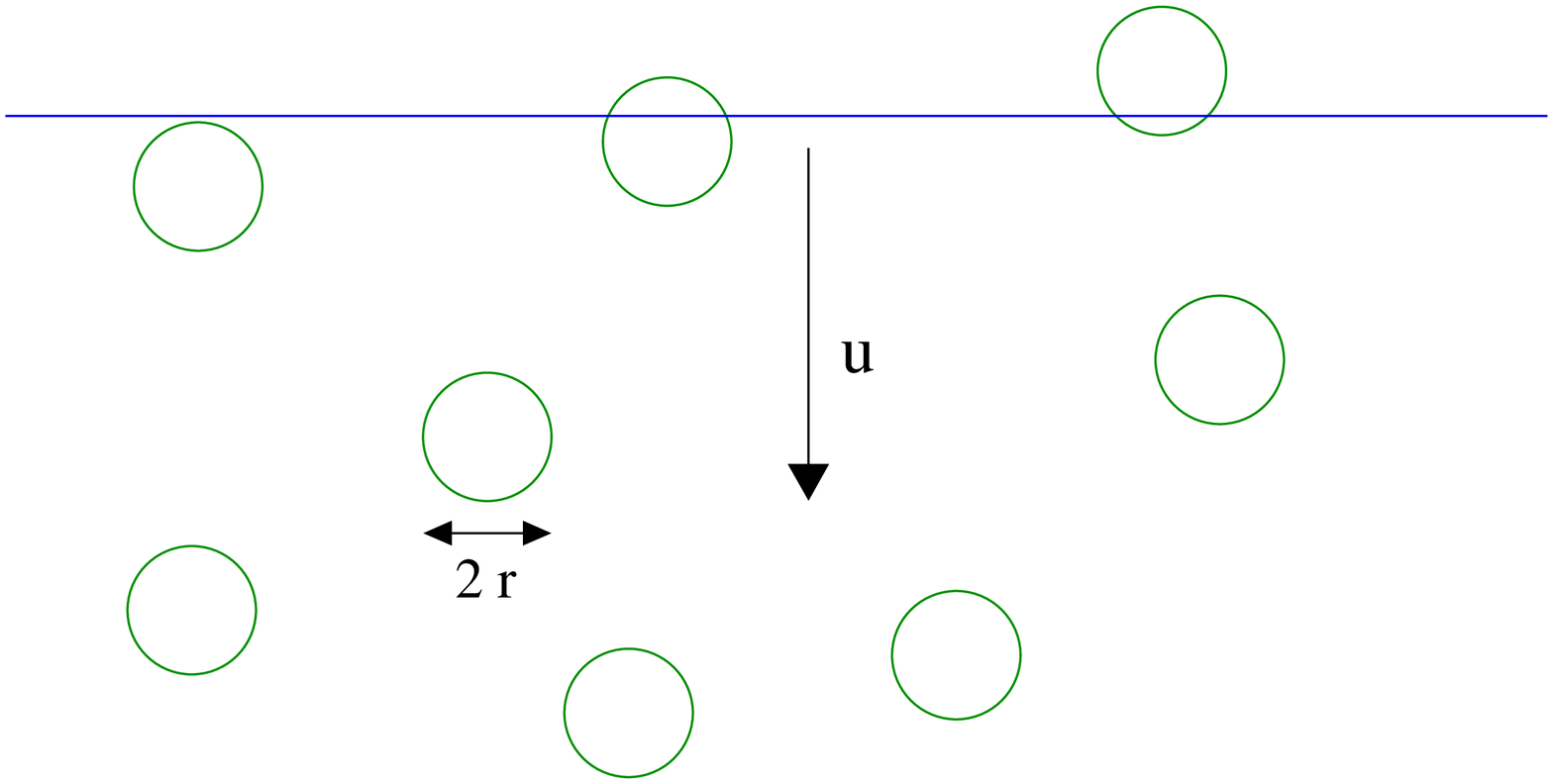}\\
  \caption{Illustration of Lineland travelling down into a higher dimensional space filled with identical static circles.} 
  $\hphantom{.}$\\ $\hphantom{.}$\\
    \label{schema2}
  \includegraphics[width=1.1\linewidth]{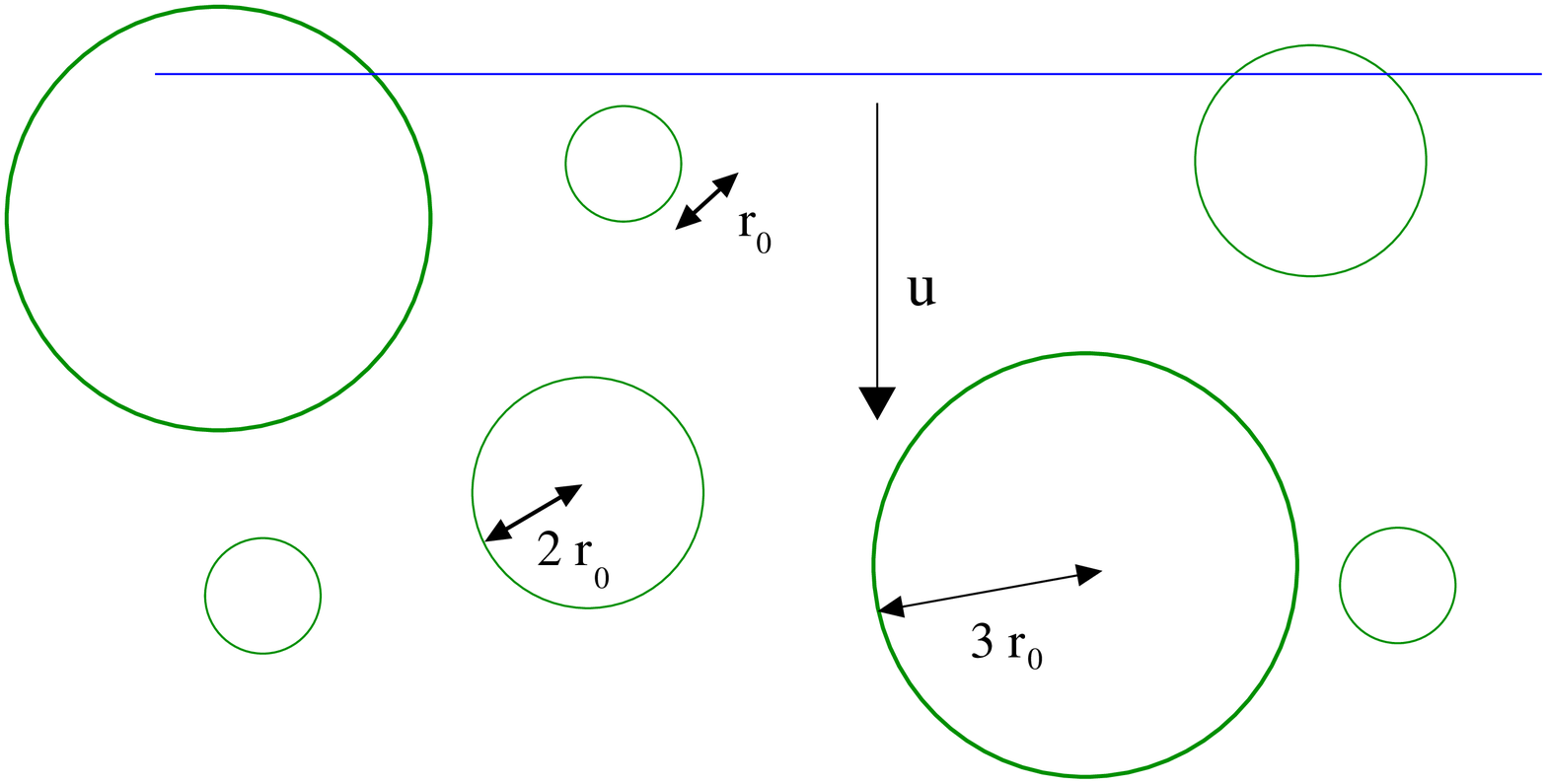}\\
  \caption{A more sophisticated scenario in which Lineland is travelling down into a higher dimensional space filled with static circles of different size.}
\end{minipage}%
\begin{minipage}{.05\textwidth}
  $\hphantom{.}$
\end{minipage}
\begin{minipage}{.4\textwidth}
  \centering
    \label{schema3}
  \includegraphics[width=1.1\linewidth]{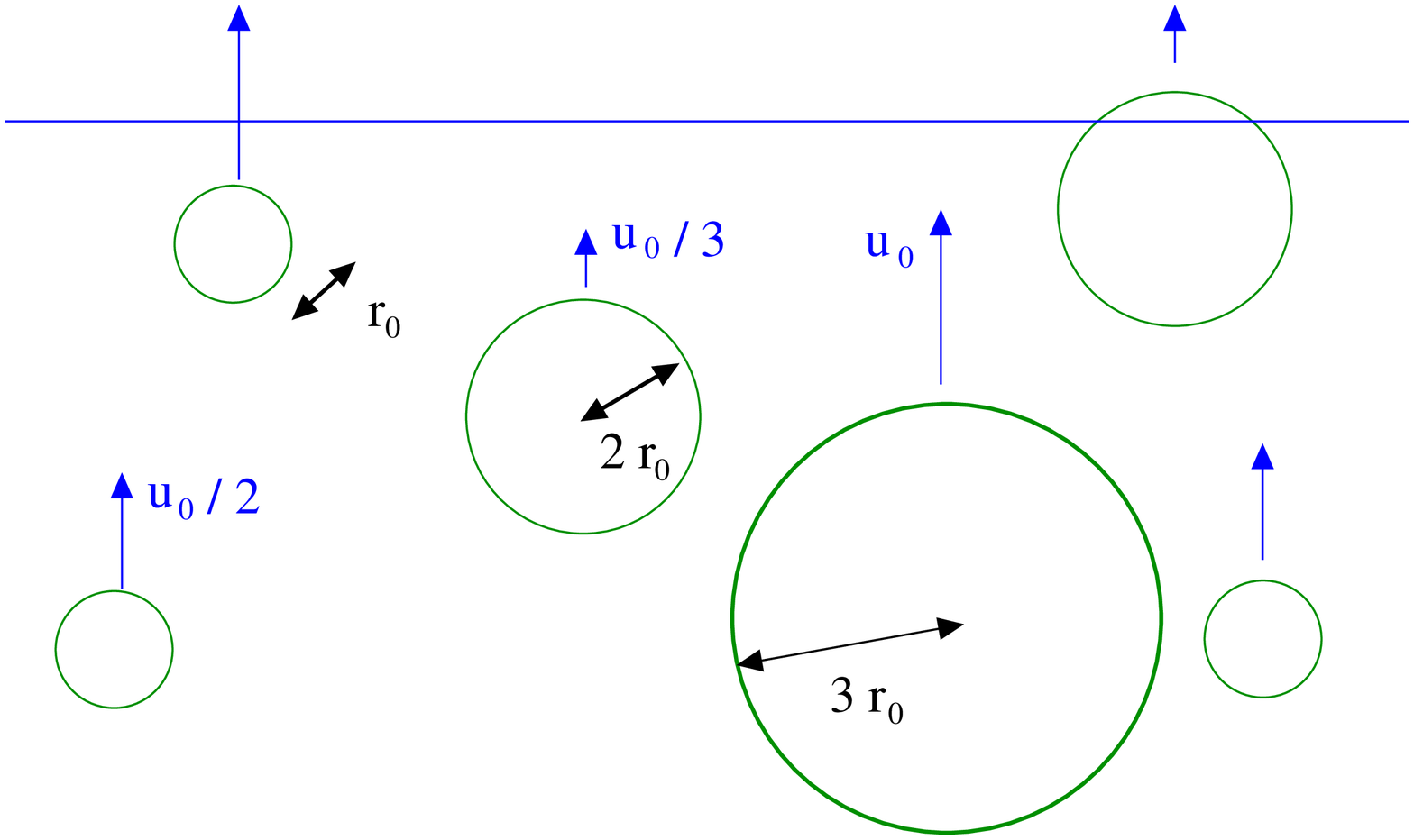}\\
  \caption{Lineland is fixed in space. Circle of different size are passing through at different speed.} 
  $\hphantom{.}$\\ $\hphantom{.}$\\
    \label{schema4}
  \includegraphics[width=0.8\linewidth]{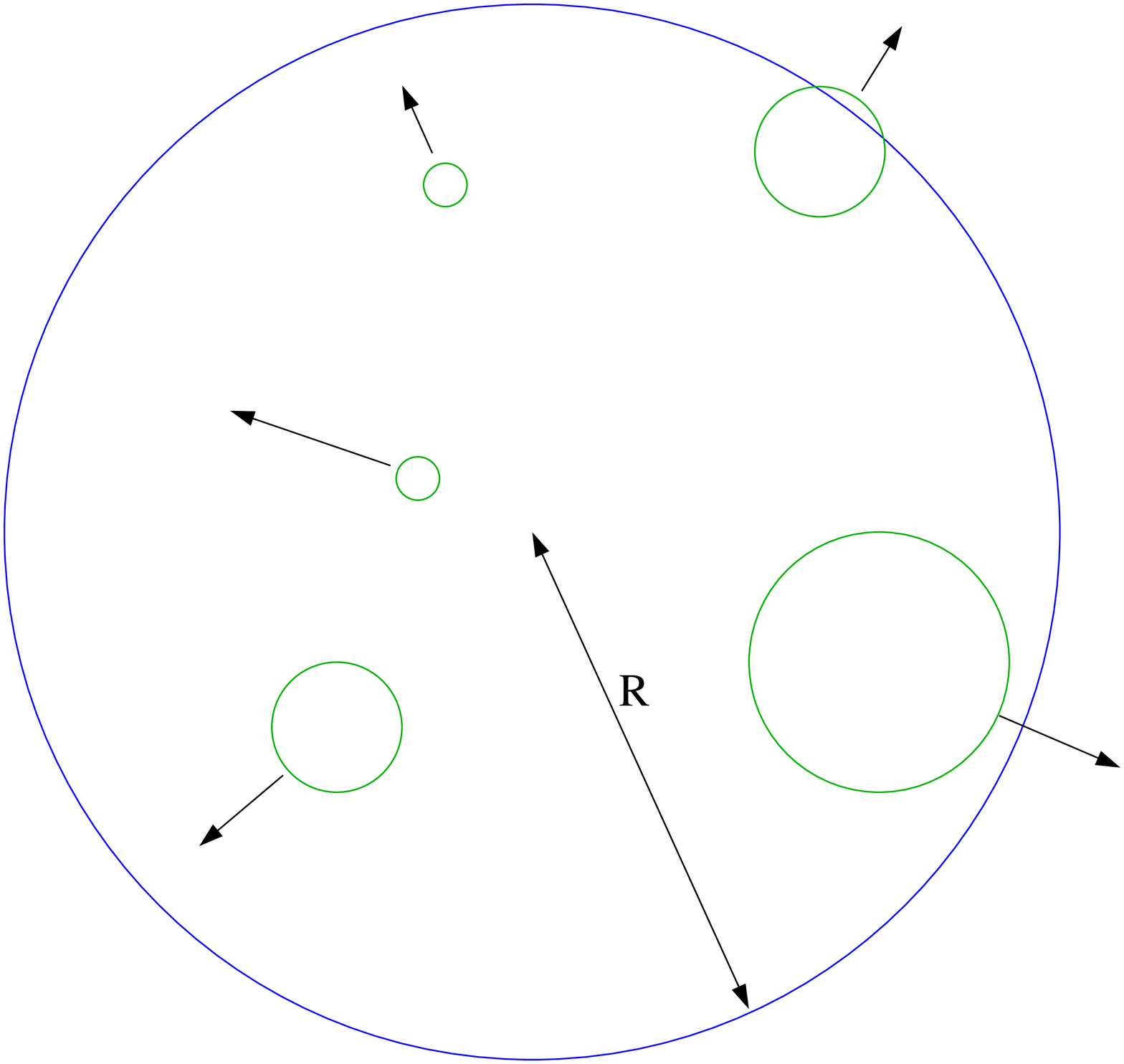}\\
  \caption{In our last scenario Lineland is a circle of radius $R$. Circles emerging from the origin are passing through with different velocities.}
\end{minipage}
\end{figure}
%%%%%%%%%%%%%%%%%%%%%%%%%
%%%%%%%%%%%%%%%%%%%%%%%%%

%%%%%%%%%%%%%%%%%%%%%%%%%
%%%%%%%%%%%%%%%%%%%%%%%%%
\begin{figure}
\centering
    \label{schema5}
  \includegraphics[width=0.9\linewidth]{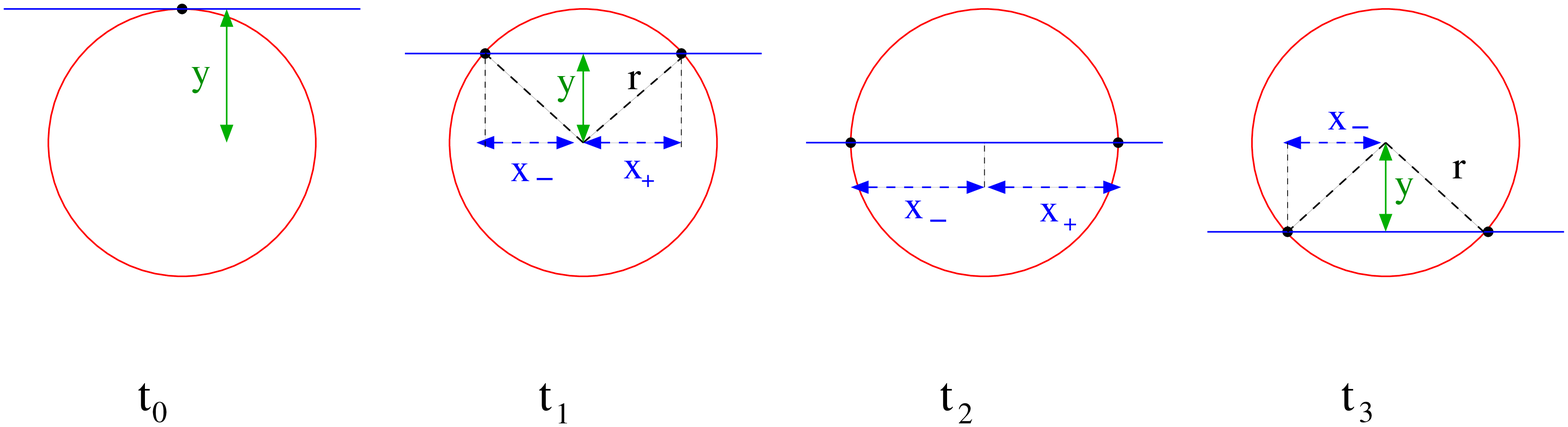}\\
  \caption{Snapshots of the interception of a circle with Lineland for different times $t_0<t_1<t_2<t_3$.}
  \label{schema_detail}
  $\hphantom{.}$\\ $\hphantom{.}$\\
\begin{minipage}{.4\textwidth}
  \centering
    \label{schema6}
  \includegraphics[width=1.1\linewidth]{xposition.eps}\\
  \caption{An ideal measurement of the position of a particle as a function of the time allows for the extraction of the constants $r$ and $u$.} 
  $\hphantom{.}$\\ $\hphantom{.}$\\
    \label{schema7}
  \includegraphics[width=1.1\linewidth]{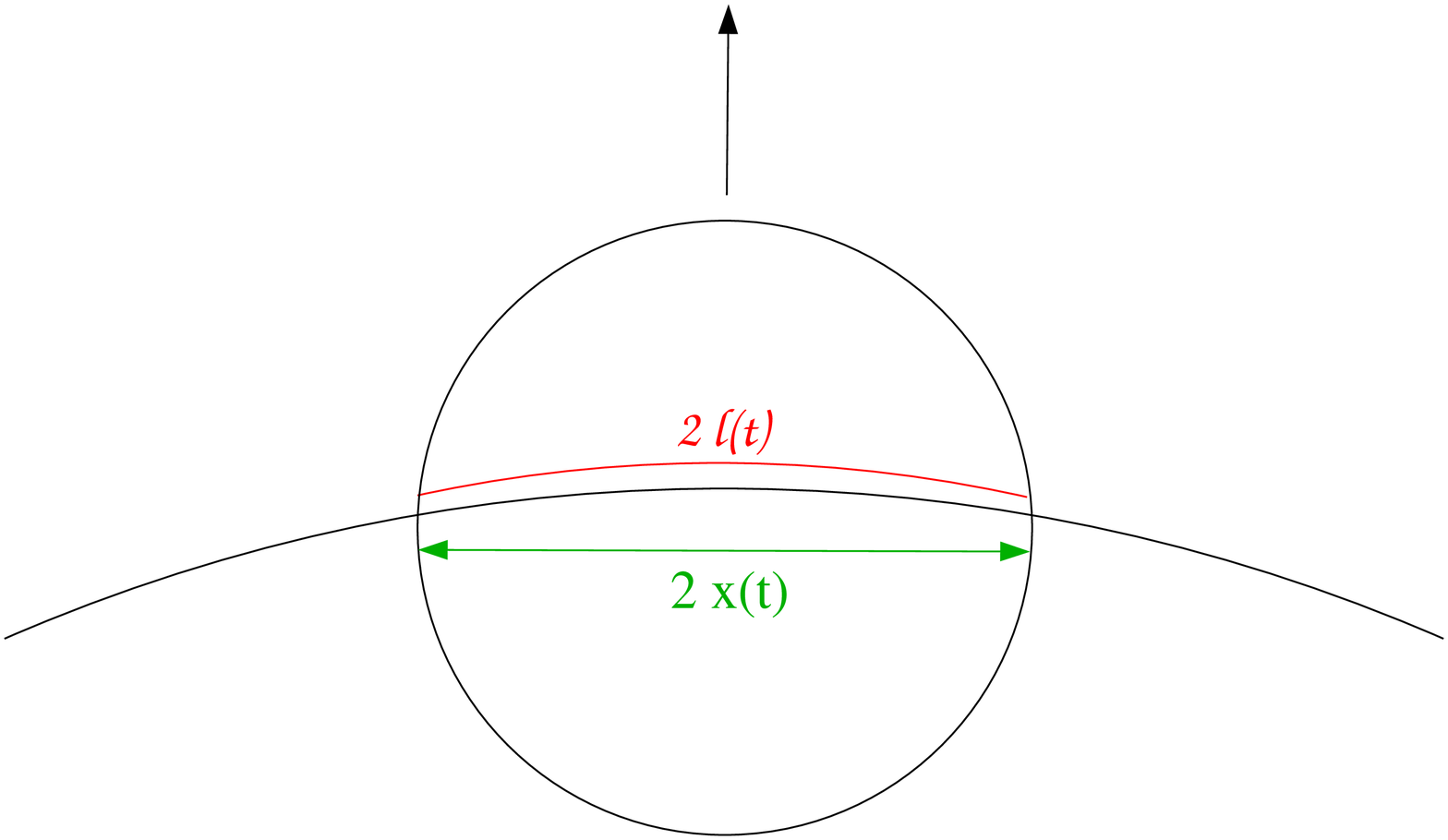}\\
  \caption{In our last scenario Lineland is a circle. Assuming $R\gg r$, the distance between particles $2l(t)$ is approximately equal to $2x(t)$.}
\end{minipage}%
\begin{minipage}{.05\textwidth}
  $\hphantom{.}$
\end{minipage}
\begin{minipage}{.4\textwidth}
  \centering
  \includegraphics[width=0.95\linewidth]{PHASESPACE.eps}\\
  \caption{Phase space trajectories described by equation ($\ref{phase_space_q}$) for effective particles of charge $q=1$ (red), $q=2$ (blue) and $q=3$ (green).}
  \label{schema8}
\end{minipage}
\end{figure}
%%%%%%%%%%%%%%%%%%%%%%%%%
%%%%%%%%%%%%%%%%%%%%%%%%%

%%%%%%%%%%%%%%%%%%%%%%%%
%%%%%%%%%%%%%%%%%%%%%%%%
\section{A Flat World in Frozen Space}
%%%%%%%%%%%%%%%%%%%%%%%%
%%%%%%%%%%%%%%%%%%%%%%%%
Let us begin by considering a line travelling at constant velocity through a two dimensional space as illustrated in figure $1$. We assume that space is filled with identical circles, randomly distributed in space,  and of radius $r$. We assume the density of circles to be small enough so that they do not overlap ($\rho\ll1$). We choose to write $y(t)=y_0-ut$ the equation of the line as it is travelling down with velocity $u$. We assume that inhabitants of Lineland can only observe events located on the line, as mentioned in the introduction. Thus as a circle passes through their world, Linelanders solely observe the intersection points with Lineland. These points which appears as particles on Lineland are the object of our attention. Initially, let us consider a circle centered on the origin, so that its equation is $x^2+y^2=r^2$. We define as $x_-$ and $x_+$ the coordinates of two intersection points, as illustrated on figure $\ref{schema_detail}$. By applying Pythagoras theorem we get the equation
\Beq\label{equation007}
x_+(t)=\sqrt{r^2-y^2(t)},
\Eeq
from which we wish to derive a "law of motion" of the form $F=ma$. Using $d x_+^2(t)/dt=2x_+(t)v_+(t)$ we successively express the velocity and acceleration as
\begin{eqnarray}
v_+(t)=u{y(t)}/{x_+(t)},\ \ \
a_+(t)=-{u^2r^2}/{x_+^3(t)}.
\end{eqnarray}
Naturally, the inhabitants of Lineland have no idea that they are observing an entirely geometric effect, but observing that the acceleration of each particle in a pair is determined by the separation, conclude that each particle exerts a force on the other which is proportional to their common mass and inversely proportional to the cube of the distance between them. Defining an arbitrary mass $m$ for every particle, we obtain the effective force
\Beq\label{Force_1}
F=-\frac{8mu^2r^2}{L^3},
\Eeq
where $L=x_+-x_-$ is the distance between particles. As Lineland is travelling through space and intercepts one of the circles, the apparent motion of the intersection points seems to be governed by the law of motion $F=m a$ where $F$ is defined by equation ($\ref{Force_1}$). The motion of the two effective particles is then easy to understand. As they appear on Lineland, with opposite (and infinite) initial velocities, the particles are travelling away from each other. They are slowly decelerating because of the attractive effective force. Eventually, they slow down to zero velocity to fall back on each other. We should mention that there is here no way to determine the mass $m$ associated to each particle. It has been arbitrary defined by Linelanders in order to impose the form of the law of motion on to the trajectory of the intersection points. This recalls to us the fact that to measure a mass, you should be able to somehow interact with the object under consideration: Linelanders cannot interact with these particles, but only observe them.

%%%%%%%%%%%%%%%%%%%%%%%%
%%%%%%%%%%%%%%%%%%%%%%%%
\subsection{Measuring slow moving particles}
Let us continue with the following scenario. We imagine what could have been the physics of these particle pairs if scientists were solely able to track slow moving particles. It implies that people of Lineland do not observe the creation of a pair, since immediately after this event the particle velocity is very large. Observations are therefore restricted to some 'large' time subsequent to the creation event. As a first approximation, we assume that the observations can only be made when the particles are near their maximum separation.
% We first start by considering circles of fixed size and show that the generalisation to circle of arbitrary size leads to the definition of an effective charge for each particles.
%\subsubsection{Fixed size circles}
%One can think that, early in the history of science on Lineland, scientist were limited in their ability to track particles. We assume that they could only manage to track slow moving particles.
In order to follow the motion of the right particle only, let us choose $y_0=0$ and $x_+(0)=0$ so that $x_+(t)=\sqrt{r^2-(ut)^2}-r$. With this choice of coordinate references we observe that $v_+(t)$ is increasing with time. People of Lineland, tracking slow particles, are here limited to "small" time $t$. At the lowest order in $t$, the equation of motion for $x_+(t)$ gives
\Beq
x_+(t)\simeq-\frac{1}{2}\frac{\left(ut\right)^2}{r}.
\Eeq
Just as we do, people of Lineland observe parabolic trajectories. At this stage however, physicists on Lineland, do not have access to the constants $r$ and $u$. To them, the equation of motion is of the form $x_+(t)=-gt^2/2$, with $g=u^2/r$. On the time scale of their experiment, the particle seems to be accelerated by a uniform field pointing to the left ($a_+=-g$). Identically, they observe the opposite trajectories for particles travelling to the right with constant acceleration $a_-=g$. It is likely, that at this stage in the history of Lineland, these observations have lead to the definition of a uniform field $g$. People of Lineland could then assume the existence of two different type of particles $+$ and $-$, which, in interaction with the field $g$, are subject to the force:
\Beq
F_{\pm}=\mp mg.
\Eeq
%%%%%%%%%%%%%%%%%%%%%%%%%%
The existence of circles of different size, as illustrated in figure $2$, would lead to different trajectories. Such observation would have been likely to lead to the definition of an extra characteristic $\sigma$: the charge of a particle. To give an example let us consider three different types of circles with radii $r=r_0$, $r=2r_0$ and $r=3r_0$. Choosing an acceleration of reference, let say: $g_0=u^2/r_0$, allows for the definition of the charge $\sigma$. For a circle of radius $r$, defining $\sigma=r_0/r$, the acceleration of the $+$ particle is
\Beq
a_+=-{u^2}/{r}=-g_0\sigma.
\Eeq
By the same argument, for the left moving particle we have $a_-=g_0\sigma$. Allowing for the definition of both negative and positive charges, the trajectories of $+$ and $-$ particles then unified under the following law
\Beq
F=-\sigma m g_0,
\Eeq
where the direction of the force is determined by the sign of the charge. Particles of negative/positive charges are respectively accelerated to the right/left. In this particular example, the charge can take six different rational values $\sigma=\pm1,\pm1/2$ and $\pm1/3$.
%%%%%%%%%%%%%%%%%%%%%%%%%%%
%%%%%%%%%%%%%%%%%%%%%%%%%%%
\subsection{Measuring fast moving particles}
We might now consider the effect of technological breakthroughs allowing the observation and measurement of fast moving particles. If capable of measuring the full trajectory of a particle, Linelanders could extract from their data, both the radius $r$ and velocity $u$ of a circle (see figure $6$). The potential energy from which equation ($\ref{Force_1}$) is derived is
\Beq
V=-\frac{4mu^2r^2}{L^2}.
\Eeq
The kinetic energy of the two interacting particles is $T={mv_-^2}/{2}+{mv_+^2}/{2}$, with $v_-=-v+=v$, leading to the total energy 
\Beq
E={mv^2}-\frac{4mu^2r^2}{L^2}.
\Eeq
In particular, for $L=2r$ the velocity vanishes so that the total energy can be expressed as $E=-mu^2$. Writing $X=L/2r$ and $Y=v/u$ the phase space trajectory of the pair is given by the equation
\Beq
X^2(Y^2+1)=1.
\Eeq
%%%%%%%%%%%%%%%%%%%%
Let us again investigate the consequences of the existence of circles of different size. As earlier this allows for the definition of the charge of the particle. We make the important observation that the total energy $E$ of a circle is independent of its radius. This time we define the charge of a particle as $q=r/r_0$ so that the potential energy can be expressed as
\Beq
V=-\frac{4|E|r_0^2q^2}{L^2}.
\Eeq
Note that $q$ is the inverse of what has been defined as the charge in the slow limit regime ($q=1/\sigma$). As done previously, we choose to consider $r=r_0$, $r=2r_0$ and $r=3r_0$, so that the charge appears to be $q=\pm1,\pm2$ and $\pm3$. Defining $X=L/2r_0$, the phase space trajectory of the system (the two particles of charge $\pm q$) takes the form
\Beq\label{phase_space_q}
X^2(Y^2+1)=q^2,
\Eeq
and the three possibilities are represented on figure $3$.
One could also choose to re-write the force as
\Beq
F=\frac{8|E|r_0^2q(-q)}{L^3}.
\Eeq
The latter equation might suggest the generalisation to interactions between particles of different charges $q$ and $q'$. To people on Lineland however, it seems that particle pairs always satisfy the constraint $q'=-q$: the underlying reason is mysterious, but they find it convenient to postulate the conservation of charge during the creation and annihilation of pairs.
%%%%%%%%%%%%%%%%%%%%%%%%%%%
%%%%%%%%%%%%%%%%%%%%%%%%%%%
%%%%%%%%%%%%%%%%%%%%%%%%%%%
%%%%%%%%%%%%%%%%%%%%%%%%%%%
\section{Lineland in an Evolving Space}
In the situation considered so far, we could equivalently regard the circles as stationary and Lineland as travelling, or Lineland as stationary and the circles as travelling. We wish now to consider the possibility that not all circles have the same speed, and so we must allow the circles to travel in space, while Lineland is kept stationary on the vertical axis. (No greater generality is achieved by having Lineland also moving.) This setup allows us to consider circles of different velocity and different size, though we retain the assumption that all circles are travelling perpendicularly to Lineland.

%%%%%%%%%%%%%%%%%%%%%%%%%%%
%%%%%%%%%%%%%%%%%%%%%%%%%%%
\subsection{A Flat World}
As mentioned above, we now choose to consider circles of different radius travelling through Lineland at different velocities. To keep things simple we consider the possible speeds as $u=u_0$, $u=2u_0$ and $u=3u_0$ and radii $r=r_0$, $r=2r_0$ and $r=3r_0$. In addition we assume that the density of circles of a given size and velocity is independent of $r$ and $u$ ($\rho(r,u)=\rho_0$). Let us extend our definition of  the charge to
\Beq
Q=\frac{u}{u_0}\frac{r}{r_0},
\Eeq
and reference energy level as $E_0=-mu_0^2$,  so that the force can be written as
\Beq
F=\frac{8|E_0|r_0^2Q(-Q)}{L^3}.
\Eeq
It follows that the possible charge values associated to a particle are given by table 1. Linelanders therefore find that there are twice as many particles of charge $2,3$ and $6$ than particles of charge $1,4$ and $9$. In other words: the density of particles of charge $Q$ is given by
\Beq
\rho(2)=\rho(3)=\rho(6)=2\rho_0, \ \ \ \rho(1)=\rho(4)=\rho(9)=\rho_0.
\Eeq
Finally, we could define an extra dimensionless quantity as $n^2={E}/{E_0}=\left({u}/{u_0}\right)^2$. Note that $n$ takes the values $n=1,2,3$. Defining $X=L/2r_0$ and $Y=v/u_0$ equation ($\ref{phase_space_q}$) becomes  $X^2(n^2+Y^2)=Q^2$. For every particle of charge $Q$ we observe that $n$ can take one of two different possible values. As inhabitants of Lineland might have done, we build table 2 and imagine how these observations would determine the law of physics taught to young physicists in Lineland.

{\it There exist $12$ different kinds of particles characterised by their charges ($Q=\pm 1,\pm2,\pm3,\pm4,\pm6,\pm9$). A particle of charge $Q$ only interacts with its partner particle of charge $-Q$ (as the charge is conserved under the creation process).  Each particle is characterised by a dimensionless number $n$. For all particles such that $\sqrt Q$ is an integer this number is unique, given by $n=\sqrt Q$. Otherwise, $n$ can takes two integer values $n_1$ and $n_2$ such that $n_1\times n_2=Q$. The density of pairs of particles $\rho(Q,n)$ being constant, given by $\rho_0$, the density of particles of charge $Q$ is given by $\rho_0$ or $2\rho_0$ depending on the properties of $Q$.}

Not taught to the young physicists is the rather peculiar notion that Lineland, the entire universe, is actually part of a larger dimensional space, and that the particles they observe are simply the intersection, with Lineland, of curved "objects" travelling in this space. Although it seems to give an explanation for some otherwise inexplicable phenomena, such as the matching of charges in particle pairs or the impossibility of measuring the particle mass, this theory is regarded as pointless metaphysics, and consideration of it restricted to departments of extremely theoretical physics.

The zoology of particles may have appeared mysterious to many Linelanders. When trying to understand table 2, a theory, which codifies the data of the table rather than explaining it, was proposed.

{\it All particles are made from combinations of $3$ elementary particles $\alpha$, $\beta$, $\gamma$ with respective charges $Q_\alpha=1$, $Q_\beta=2$ and $Q_\gamma=3$. It is only possible to group identical particles together, combining particles of different types is forbidden (for unknown reasons). It follows that a particle of charge $Q=2$, could be the particle $\beta$ or the combination of $2\alpha$. Identically, a particle of charge $Q=6$ could be $3\beta$ or $2\gamma$. Note that the dimensionless number $n$ is the number of elementary particle needed to compose the new one. Curiously, since $E=n^2E_0$, the energy of a composed particle grows with the square of the number of elementary particles which composed it.}
%%%%%%%%%%%%%%%%%%%
\begin{table}[h!]
\centering
   \begin{tabular}{| c || c | c | c |}
   \hline
     	&  $r_0$ 	& $2r_0$ & $3r_0$ \\ \hline	\hline
    $u_0$ &  $1$ 	& $2$ & $3$ \\ \hline
    $2u_0$ &  $2$ 	& $4$ & $6$ \\ \hline
    $3u_0$ 	&  $3$ 	& $6$ & $9$ \\ \hline
  \end{tabular}
  \caption{This table presents the charge $Q$ obtained by considering circles of radii $r=r_0$, $2r_0$, $3r_0$ and velocities $u=u_0$, $2u_0$ and $3u_0$.}
\end{table}
\begin{table}[h!]
\centering
   \begin{tabular}{ | c || c |}
	\hline
     Charge ($Q$)	&  Dimensionless number $n$ 	\\ \hline \hline
    $1$ &  $1$ 	\\ \hline 
    $2$ &  $1$ or $2$	\\ \hline
    $3$ &  $1$ or $3$\\ \hline
    $4$ &  $2$\\ \hline
    $6$ &  $2$ or $3$\\ \hline
    $9$ &  $3$ \\ \hline
  \end{tabular}
  \caption{This table present the possible dimensionless number ($n$) associated to particles of charge $Q$.}

\end{table}

%%%%%%%%%%%%%%%%%%%%%%%%%%%
%%%%%%%%%%%%%%%%%%%%%%%%%%%
%%%%%%%%%%%%%%%%%%%%%%%%%%%
%%%%%%%%%%%%%%%%%%%%%%%%%%%
\subsection{A Not so Flat World}
We now wish to consider one last, more general scenario which will leads scientists on Lineland to reject the theory they have build so far. Just like us, even the most theoretical physicist Linelanders worked with a wrong idea for years. All their observations were consistent with Lineland being a straight line, while in fact Lineland is not flat but curved. The observations were simply insufficiently accurate to show this. In fact, Lineland is not Lineland at all, but Circleland, a circle of radius $R\gg1$. We can now return to our initial calculation to see what correction will follow from this consideration. Let us start by writing $x^2+y^2=R^2$ the equation of Lineland (now Circleland). We can think of the small circles as emerging from the origin and travelling radially. We write as $x^2+[y-h(t)]^2=r^2$ the equation of a small circle with $h(t)=ut$. As illustrated on figure $\ref{schema7}$, the distance between the two intersection points is $L(t)=2l(t)$. We should point out that $l(t)$ is only approximately equal to $x(t)$. Under this approximation ($l(t)\simeq x(t)$), which is valid if $R \gg r$, we have
\BeqA
\label{eq1}
l^2(t)&=&R^2-y^2,\\
l^2(t)&=&r^2-[y-h(t)]^2.
\label{eq2}
\EeqA
The first equation gives $d y/dt=- (d l^2/dt)/2y$, while Eq. ($\ref{eq2}$) leads to 
\Beq
v(t)=\frac{u}{l(t)}\frac{\sqrt{r^2-l^2}}{1-\sqrt{z}},\ \ \ z=\frac{r^2-l^2}{R^2-l^2}.
\Eeq
After some work, one can finally show that the acceleration is given by
\Beq
a(t)=
-\frac{u^2r^2}{l^3}
\frac{1}{(1-\sqrt{z})^2}
\left(
1
+
\frac{l^2}{r^2}\left(1-z\right)\frac{\sqrt{z}}{1-\sqrt{z}}
\right).
\Eeq
When $R\gg r$, one has $z\ll1$ so that to a good degree of approximation the previous equation becomes
\Beq
a(t)=
-\frac{u^2r^2}{l^3}
\left[
1
+\frac{r}{R}\left(2+\frac{l^2}{r^2}\right)\sqrt{1-\frac{l^2}{r^2}}
\right].
\Eeq
It follows that the force is 
\Beq
F=
-\frac{8mu^2r^2}{L^3}
\left[
1
+\frac{r}{R}\left(2+\frac{L^2}{(2r)^2}\right)\sqrt{1-\frac{L^2}{(2r)^2}}
\right].
\Eeq
Sufficiently accurate measurements of the particle trajectories, following further technological advances, show that the force deviates slightly from the inverse cube law which has been accepted as fact since the observations of fast particles became possible. Experimentally, the radius $R$ could now be extracted by measuring variations from the previous law. Using equations ($\ref{eq1}$) and ($\ref{eq2}$) we show that
\Beq
l(t)=\sqrt{R^2-\frac{1}{4}\left(h(t)+\frac{R^2-r^2}{h(t)}\right)^2}.
\Eeq
At the lowest order in $1/R$, deviations from equation ($\ref{equation007}$) are given by
\Beq
l^2(t)-x^2(t)\simeq\frac{(ut)^3}{R}\left[1-\left(\frac{r}{ut}\right)^2\right].
\label{eq22}
\Eeq
Note that equation ($\ref{eq22}$) is an odd function, reflecting the fact that, if the trajectory of a particle was symmetric in our first scenario, the symmetry is now broken by having Lineland curved. It follows that the trajectories associated to the deceleration and acceleration of a particle are not identical. And, as equation ($\ref{eq22}$) indicates, deviations from the previous results are exactly anti-symmetric.

As the theory is being modified and adjusted to fit experimental results, it loses its elegance while its complexity increases drastically. Science on Lineland was then in need for an alternative theory. As a consequence, the very theoretical physicists find themselves in a position to explain the more accurate force law by the mean of a geometrisation of physics. Embedding their world in a higher dimensional space where particles appears as the interseption points with circles gives a natural explanation to charge conservation and even allow for predictions of higher order corrections by using a more accurate expression for $l(t)$. But at this point we have to leave the development of physics in Lineland, while the scientists await the ability to test the still more accurate prediction which will enable them to conclude beyond reasonable doubt that their universe is, indeed, embedded in a higher dimensional one.

%%%%%%%%%%%%%%%%%%%%%%%%%%%%%
%%%%%%%%%%%%%%%%%%%%%%%%%%%%%
\section{The Inverse Problem}
The inverse problem, associated to the story of Lineland, arises quite naturally. For simplicity purposes, we will assume Lineland to be flat ($R\rightarrow\infty$). We wish to conclude this paper by looking for the geometrical shapes which, when crossing Lineland, lead to a given effective force law. To solve this problem, we start by writing $\lambda(x)$ as the equation of the curve of the object under consideration. As it is travelling up through Lineland at constant velocity its full equation is $y=\lambda(x)+ut$. The interceptions points $x(t)$ are therefore solution of $\lambda(x(t))+ut=0$. By differentiation it follows that
\Beq\label{eq23}
v(t)\lambda'(x(t))+u=0,
\Eeq
which leads to the differential equation $mu^2\lambda''(x)+F(x)(\lambda'(x))^3=0$. At this stage, we could directly solve the latter differential equation. We however choose to consider symmetrical shapes so that $x_+(t)=-x_-(t)$ and $v_+(t)=-v_-(t)$ for all time. It follows that the kinetic energy is $T=E-V(2x)=mv^2$ so that Eq. ($\ref{eq23}$) becomes
\Beq
\lambda'(x)=\mp\frac{u\sqrt{m}}{\sqrt{E-V(2x)}},
\label{eq24}
\Eeq
where the sign is function of the direction of the velocity of the particle. When considering $V(x)=-4|E|r^2/x^2$ with positive or negative total energy ($E=\pm m u^2$) the integration of equation ($\ref{eq24}$) leads to either the equation of a circle or the equation of an hyperbola
\Beq
\lambda^2(x)\mp x^2=r^2.
\Eeq
Trivially, considering the quadratic potential $V(x)=|E|x^2/4r^2$ (with $E=mu^2$) leads to the trigonometric function
$\lambda(x)=r\arcsin(x/r)+k$, where $k$ is some constant of integration. An illustration of the corresponding geometrical shapes is presented in figure $10$. We finally choose to generalised the potential energy to $V(x)=-|E|(2r)^{2n}/x^{2n}$ (with $E=-mu^2$). Integration of equation ($\ref{eq24}$) leads to the hypergeometric function
\Beq
\lambda_n(x)=g_n(x) \hphantom{.}_2F_1\left(\frac{1}{2},\frac{n+1}{2n},\frac{3n+1}{2n},\left(\frac{x}{r}\right)^{2n}\right)+ k,\ \ \ g_n(x,r)=\frac{x}{n+1}\left(\frac{x}{r}\right)^n.
\label{eqYY}
\Eeq
Similarly, generalisation of the harmonic potential to the form $V(x)=|E|x^{2n}/(2r)^n$ leads to 
\Beq
\lambda_n(x)=g_0(x,r) \hphantom{.}_2F_1\left(\frac{1}{2},\frac{1}{2n},\frac{2n+1}{2n},\left(\frac{x}{r}\right)^{2n}\right)+ k.
\label{eqZZ}
\Eeq
Figures $9$ and $10$ present the geometrical shapes associated to equations ($\ref{eqYY}$) and ($\ref{eqZZ}$). We note that the shapes generated from $V(x)\propto-1/x^{2n}$ are smooth, indicating that the velocities of the effective particles at creation and annihilation are infinite. On the other side, the curve obtained for $V(x)\propto x^{2n}$ are not differentiable at $x=0$, indicating that the initial and final velocities of the particles are finite. 

%%%%%%%%%%%%%%%%%%%%%%%%%
%%%%%%%%%%%%%%%%%%%%%%%%%
\begin{figure}
\centering
\begin{minipage}{.4\textwidth}
  \centering
    \label{WalkingtheLine}
  \includegraphics[width=1.1\linewidth]{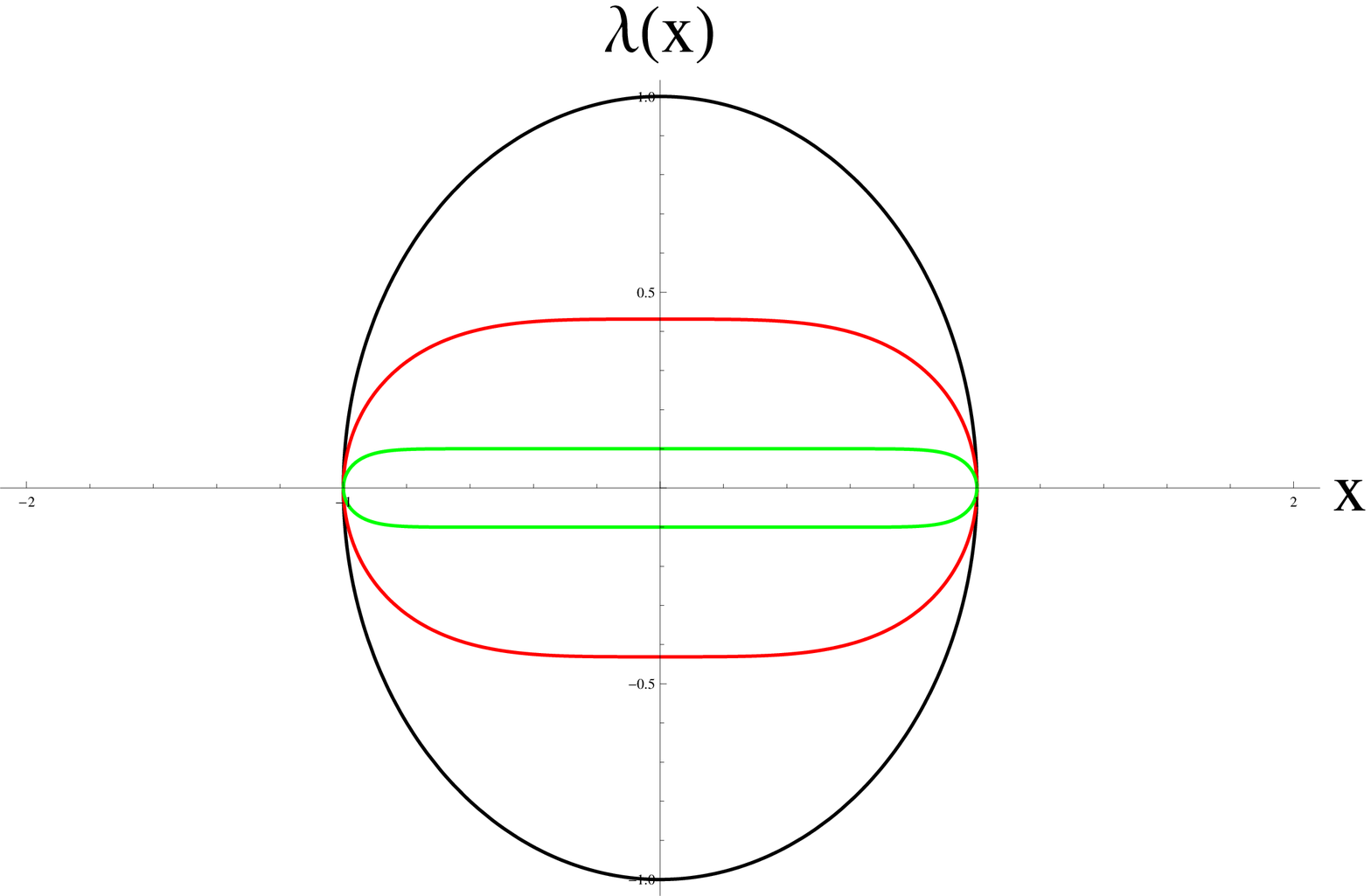}\\
  \caption{Geometrical shapes associated to a generalised potential $V(x)\propto -1/x^{2n}$ for $n=1$ (black), $3$ (red) and $15$ (green)} 
\end{minipage}%
\begin{minipage}{.05\textwidth}
  $\hphantom{.}$
\end{minipage}
\begin{minipage}{.4\textwidth}
  \centering
    \label{WalkingtheLine2}
  \includegraphics[width=1.1\linewidth]{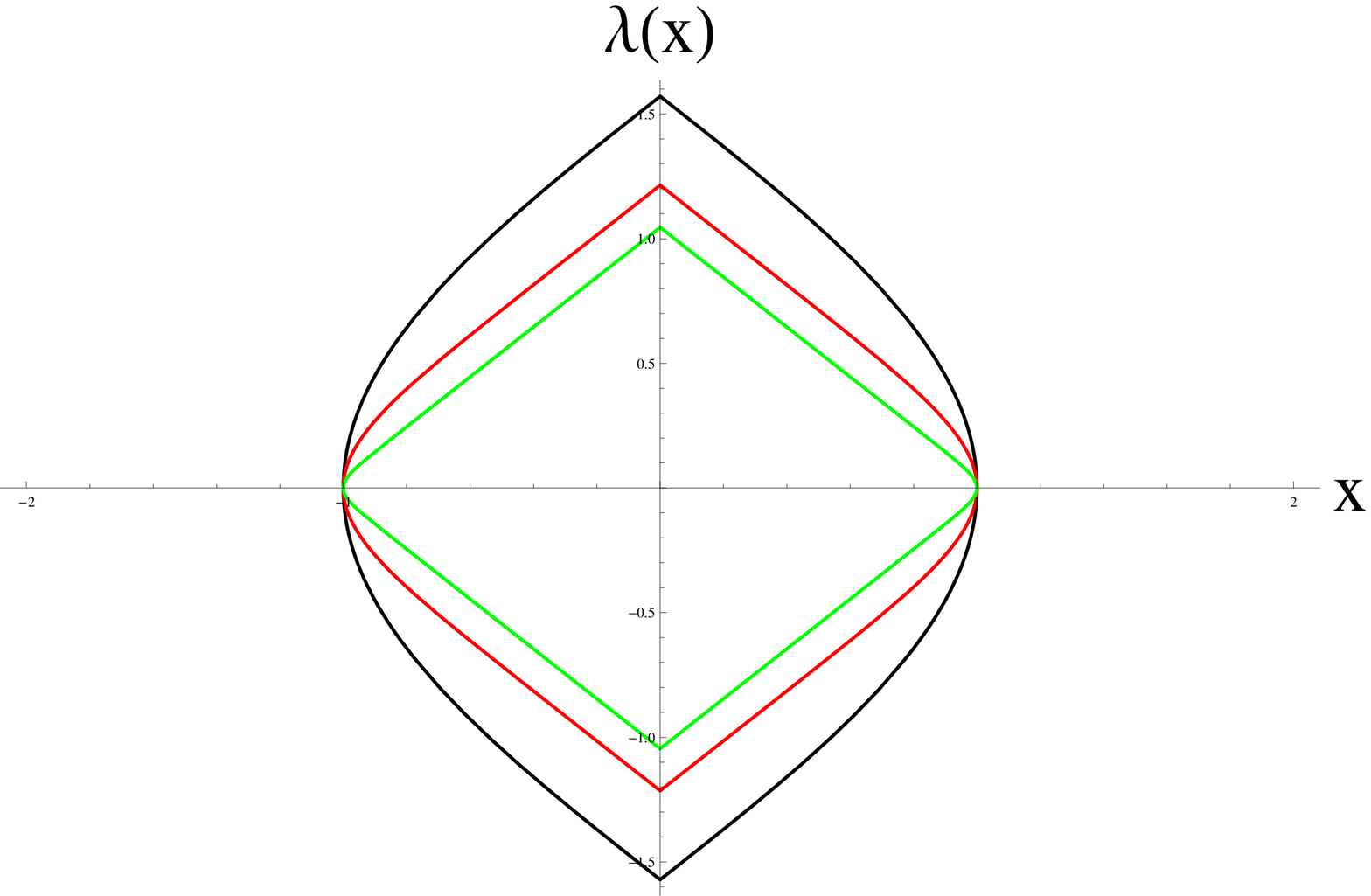}\\
  \caption{Geometrical shapes associated to a generalised potential $V(x)\propto x^{2n}$ for $n=1$ (black), $3$ (red) and $15$ (green)} 
\end{minipage}
\end{figure}
%%%%%%%%%%%%%%%%%%%%%%%%%
%%%%%%%%%%%%%%%%%%%%%%%%%

%%%%%%%%%%%%%%%%%%%%%%%%%%%%%
%%%%%%%%%%%%%%%%%%%%%%%%%%%%%
\section{Summary}
Considering a one dimensional universe, we have presented aspects of the effective theory describing the dynamics of points which are intersections with geometrical shapes living in a higher dimensional space in which Lineland is embedded. Starting by considering circles of fixed radius and travelling at a given velocity, we successively relax these constraints. The physics emerging on Lineland describes the creation and annihilation of particles as governed by an attractive force $F\propto1/L^3$ ($L$ being the distance between particles). Interestingly, when restraining our study to particles travelling at low velocity, the effective force appears to be constant $F=\pm g$, leading to parabolic trajectories. The introduction of circles of different sizes, travelling at different velocities leads to the definition of two effective particle characteristics: the charge $Q=ru/u_0r_0$ and a dimensionless number $n=u/u_0$. The phase space trajectories can then be written as $X^2(n^2+Y^2)=Q^2$. In a last scenario, we reveal that Lineland is in fact a circle of radius $R$. From its centre, circles of smaller size ($r\ll R$) are emerging and travelling at constant velocity in a straight line. This last setup breaks the creation/annihilation  symmetry so that particles in the creation and annihilation phase do not quite follow the same path. Corrections to the original effective law are then calculated at the first order in $1/R$. The theory drastically loses its elegance and therefore has to be discard. Following Occam's razor a geometrical description of the observed phenomenon should recommend itself to Linelanders. In conclusion, we addressed the inverse problem by giving alternative shapes which would lead to various effective potentials of the form $V(x)\propto -1/x^{2n}$ and $V(x)\propto x^{2n}$.

%%%%%%%%%%%%%%%%%%%%%%%%%%%%%
%%%%%%%%%%%%%%%%%%%%%%%%%%%%%

%%%%%%%%%%%%%%%%%%
%%%%%%%%%%%%%%%%%%
%%%%%%%%%%%%%%%%%%
%%%%%%%%%%%%%%%%%%
%%%%%%%%%%%%%%%%%%
%%%%%%%%%%%%%%%%%%
%%%%%%%%%%%%%%%%%%
%%%%%%%%%%%%%%%%%%
\section*{Acknowledgements}
%\ack
The authors would like to thanks all AMRC members for their support and constant efforts to protect curiosity driven research. In particular T. Platini extends his acknowledgements to the group of statistical physics in Nancy.

\end{document}